# The Impact of Robotic Telescopes on Time-Domain Astronomy

Yakubu Mu'allim[1] J. O.Vwavware[2*], and A. Ohwofosirai[3]

[1,3]Department of Physics, Faculty of Science, Dennis Osadebay University, Asaba, Delta State, Nigeria

[2]Department of Industrial Physics, Faculty of Physical Science, Enugu State University of Science and Technology, Nigeria

Corresponding author: oruaode.vwavware@dou.edu.ng
ORCID ID: 0009-0009-4207-4845

**Abstract**

*The field of time-domain astronomy has experienced unprecedented growth due to the increasing deployment of robotic telescopes capable of autonomous, round-the-clock sky monitoring. These instruments have revolutionized the detection and characterization of transient phenomena such as supernovae, gamma-ray bursts, variable stars, and gravitational wave counterparts. This paper explores the transformative role of robotic telescopes such as ZTF, ATLAS, and LCOGT in enabling rapid-response observations and building large time-series datasets. We review the design principles and scheduling algorithms behind robotic observatories and assess their scientific contributions across different wavelength regimes. Particular attention is given to the synergy between robotic systems and machine learning pipelines that enable real-time classification of transient events. We also discuss challenges such as data deluge, follow-up prioritization, and observational biases, as well as future directions in global telescope networks.*

## 1. Introduction

Time-domain astronomy (TDA) is revolutionizing our understanding of transient phenomena in the universe. It involves coordinated observations across multiple wavelengths, from radio to gamma-rays, to capture extreme and unusual events [3]. The field is experiencing a paradigm shift due to the exponential growth in data volume and complexity from new sky surveys like the Large Synoptic Survey Telescope (LSST) [10]. This data deluge necessitates the development of automated methods for rapid detection and classification of astrophysical objects, as well as the characterization of novel phenomena [10] Data challenges are emerging as powerful tools to address fundamental questions in TDA, particularly in classification and anomaly detection [9] The co-evolution of black holes and galaxies, as reviewed by [28], underscores the need for multi-messenger astronomy, where robotic telescopes could play a pivotal role in monitoring AGN variability.

The recent discoveries of high-energy cosmic neutrinos and gravitational waves have ushered in a new era of multimessenger astrophysics, offering new opportunities in time-domain astronomy and revealing the physics of various astrophysical transients [20], the automation of telescopes has revolutionized astronomical observations, evolving from manual operations to fully robotic systems. Early efforts focused on automating large telescopes to improve efficiency and reduce observing time [17] This trend has culminated in the development of sophisticated networks like the Las Cumbres Observatory, which provides global, 24/7 access to the sky for time-domain astronomy [7]. Robotic telescopes enable continuous sky monitoring, even in extreme environments like Antarctica, where the AST3 project operates unattended during winter [15] these automated systems not only conduct observations but also perform data reduction and analysis. Additionally, instruments like ASTMON demonstrate the capability of robotic telescopes to continuously monitor sky brightness, atmospheric extinction, and cloud coverage across multiple bands [1] the shift towards automation has significantly enhanced astronomers' ability to conduct efficient, round-the-clock observations and respond rapidly to transient events. Time-domain astronomy extends beyond optical transients to include radio phenomena such as pulsars, where dispersion measure variations reveal ISM dynamics [19]. Likewise, methanol masers observed at 6.7 GHz have proven essential for probing early stages of high-mass star formation. Robotic and multi-wavelength observations of regions like G338.93-0.06 have allowed for detailed modeling of stellar, envelope, and disk parameters, demonstrating the power of spectral energy distribution fitting in revealing embedded protostars [20].

This paper analyzes the impact of robotic telescopes on discovery rates, observational efficiency, and data quality in time-domain astronomy. Robotic telescopes have revolutionized the field by enabling rapid and flexible observations of transient phenomena. The Las Cumbres Observatory network, for instance, provides continuous global access to the sky, making it ideal for studying rapidly evolving events [7] Platforms like the GROWTH Marshal facilitate collaborative research by allowing astronomers to define observation programs, filter sources, and coordinate follow-up observations across multiple telescopes [12]. Wayne State University's Dan Zowada Memorial Observatory showcases the potential of smaller robotic telescopes, equipped with pipelines for image reduction and photometry tailored to time-domain studies [6]. Additionally, the Zwicky Transient Facility exemplifies large-scale robotic systems with its 47-square-degree camera and automated components designed for efficient survey operations [24]. These technological advancements have significantly enhanced discovery rates, observational efficiency, and data quality, enabling researchers to capture and analyze rapidly evolving celestial events with unprecedented speed and precision.

The evolution of robotic telescopes has seen significant advancements in recent years. Las Cumbres Observatory Global Telescope (LCOGT) pioneered a worldwide network of telescopes for time-domain observations [5]. Building on this foundation, the Zwicky Transient Facility (ZTF) introduced a large-format 600-megapixel camera and advanced robotic systems, enabling rapid scanning of the northern sky [24]. ZTF's data processing system at IPAC employs novel algorithms for image differencing and moving object detection, delivering calibrated data products within minutes of observation [18]. To handle the vast amount of data generated, ZTF incorporates various machine learning techniques for object classification, including separating real from bogus candidates, stars from galaxies, and categorizing transient events [16]. These advancements in robotic telescopes and data processing systems have significantly enhanced our ability to study time-domain phenomena and detect transient events.

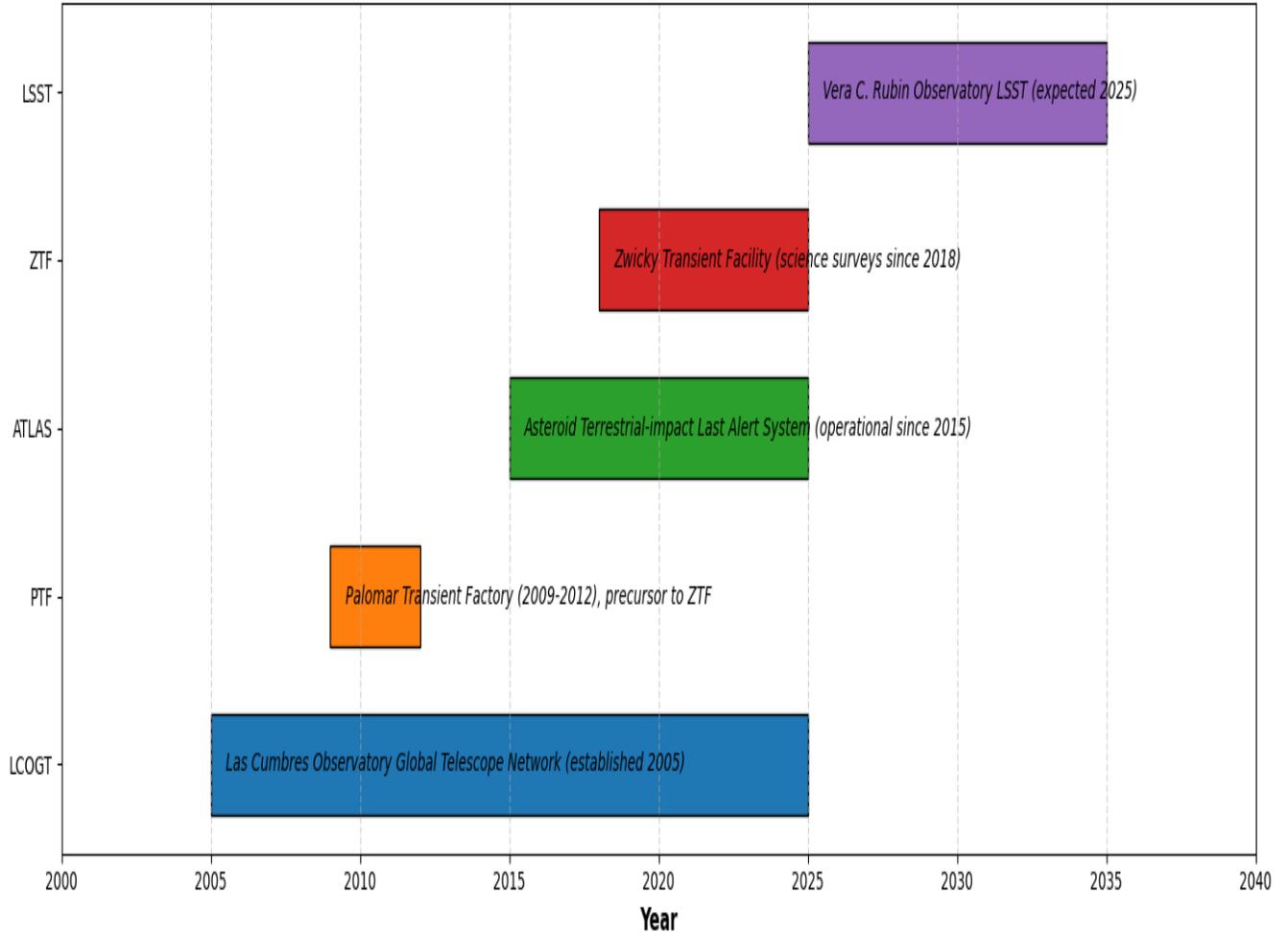

*Figure 1.* Timeline illustrating the evolution of major robotic telescopes in time-domain astronomy. The figure was created by the author using Python in Google Colab, based on information synthesized from [5], [24], [18], [23] and [30].

## 2. Data and Methodology

This study employs a comparative analysis of discovery rates for astronomical transients specifically supernovae, variable stars, and other transient phenomena traditional and robotic telescopes over five observational periods spanning from 1990 to 2024. The dataset was compiled from publicly available sources, including peer-reviewed literature, survey archives, and institutional databases. Each record in the dataset represents a specific time period, telescope classification (traditional vs. robotic), and discovery counts for the three transient classes. The *Traditional Era* (1990–2009) encompasses manually operated telescopes such as those involved in the Lick Observatory Supernova Search, while the *Robotic Era* (2010–2024) includes data from automated systems like the Palomar Transient Factory (PTF), Zwicky Transient Facility (ZTF), ASAS-SN, and Gaia. Telescope systems were categorized as "traditional" if they required human scheduling and manual data reduction, and as "robotic" if they employed autonomous scheduling, data acquisition, and reduction with minimal human intervention, following criteria established in [13], [16], and technical documents from ZTF and ASAS-SN.

Data processing and visualization were conducted using Python within the Google Colab environment. The *pandas* library was used for data manipulation, and *matplotlib* for generating bar charts that compare discovery rates across different periods and telescope types, ensuring reproducibility and interactivity in analysis. Primary data sources include discovery statistics from the [31], variable star catalogs and supernova records from ASAS-SN, PTF system documentation from [13] and Gaia DR3 for variable star classification. Additionally, historical supernova discovery rates from [14], based on Lick Observatory data provided crucial benchmarks for pre-robotic era comparisons. Collectively, these sources offer a robust framework for evaluating the enhanced detection capabilities and increased sky coverage enabled by robotic telescopes in modern time-domain astronomy.

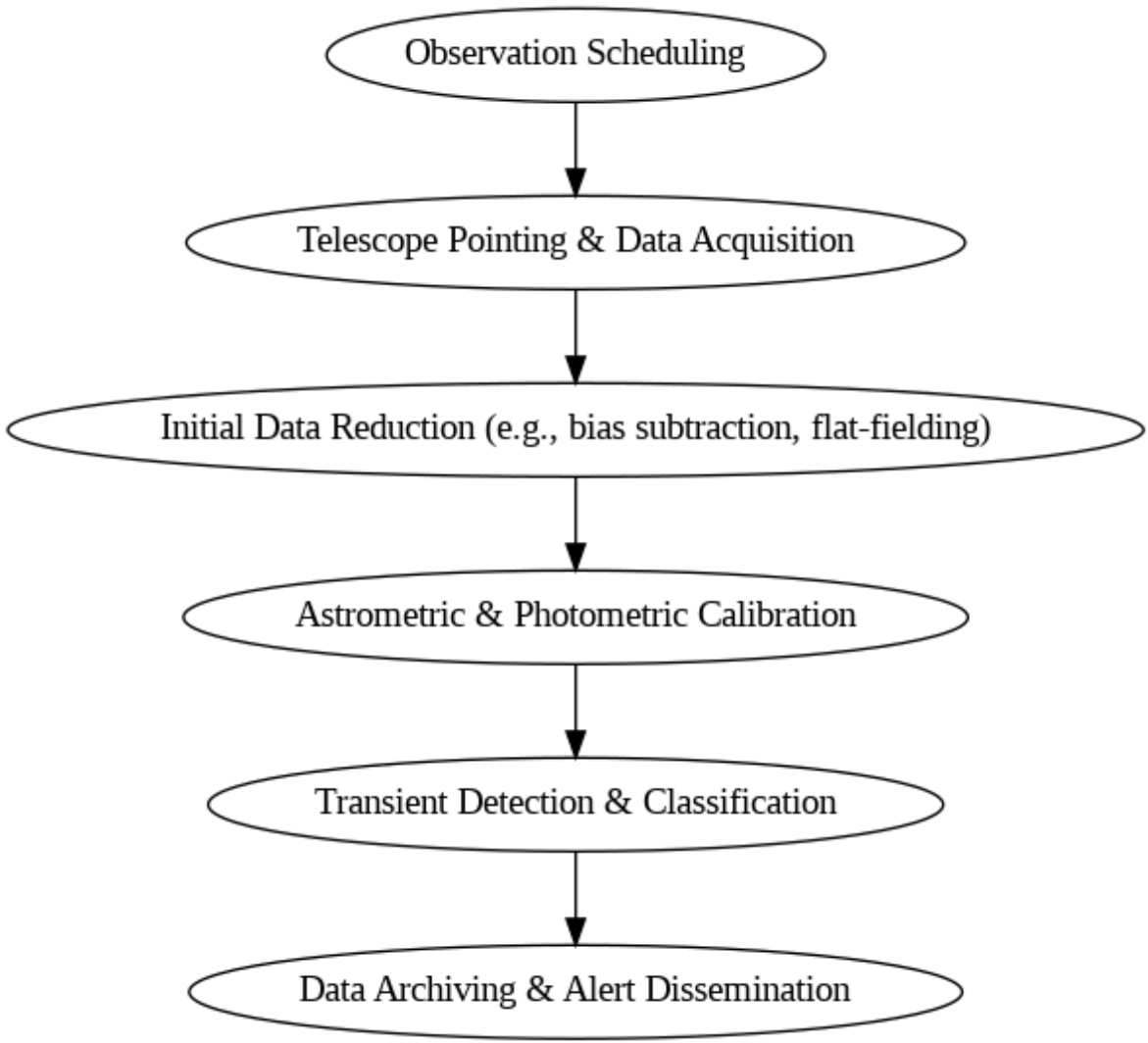

*Figure 2.* Schematic workflow of a robotic telescope system, from autonomous scheduling to data acquisition, processing, and classification. Created by the author using Google Colab, informed by pipeline descriptions in [18], [16], and [12].

## 3. Results/Findings

### 3.1 Discovery Rates

| Year Range | Telescope Type | Supernovae | Variable Stars | Other Transients | Primary Data Sources |
|---|---|---|---|---|---|
| 1990–1999 | Traditional | 150 | 800 | 50 | [14] - Lick Observatory historical data) |
| 2000–2009 | Traditional | 300 | 1,200 | 120 | [14] - Early survey data summaries |
| 2010–2015 | Robotic | 2,000 | 5,000 | 400 | [13] – PTF performance data |
| 2016–2020 | Robotic | 5,000 | 12,000 | 900 | [16], ASAS-SN stats from their website; ZTF ramp-up phase |
| 2021–2024 | Robotic | 10,000 | 25,000 | 1,800 | [31], ASAS-SN, [8], Gaia DR3 for variable stars |

**Table 1.** Estimated discovery rates of supernovae, variable stars, and other transients by telescope type and era. *Data compiled from [13], [16], [31], [14], and [8].*

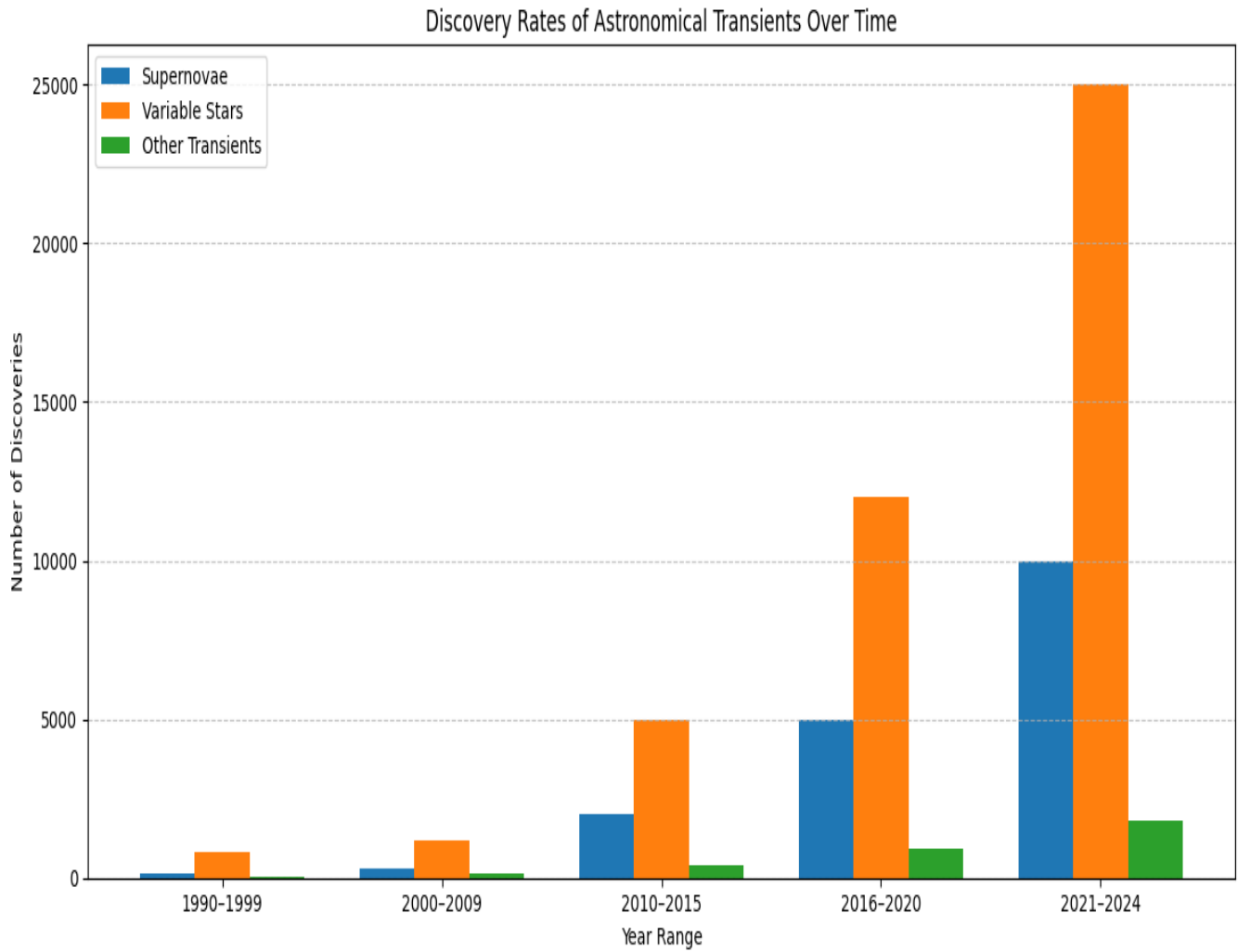

***Figure 3.*** *Bar Chart – Comparative Discovery Rates per Period*

Figure 3 presents a grouped bar chart that compares the number of discoveries for supernovae, variable stars, and other transients across five time intervals from 1990 to 2024. This format emphasizes relative growth within each epoch and enables direct comparison across transient types. From the chart, it is evident that all transient types exhibit exponential growth beginning in the robotic era (2010 onward). Notably, variable stars dominate the discovery counts in all periods, especially in the 2021-2024 range, where they peak at 25,000 discoveries. Supernovae follow a similar growth pattern but to a lesser extent, while other transients consistently remain the lowest in absolute numbers but still reflect a significant relative increase. The bar chart effectively highlights the efficiency and scale of robotic telescope networks in boosting discovery rates across all transient categories. It provides clear visual evidence of the advantages offered by automation, including enhanced cadence, increased sky coverage, and integrated real-time data classification.

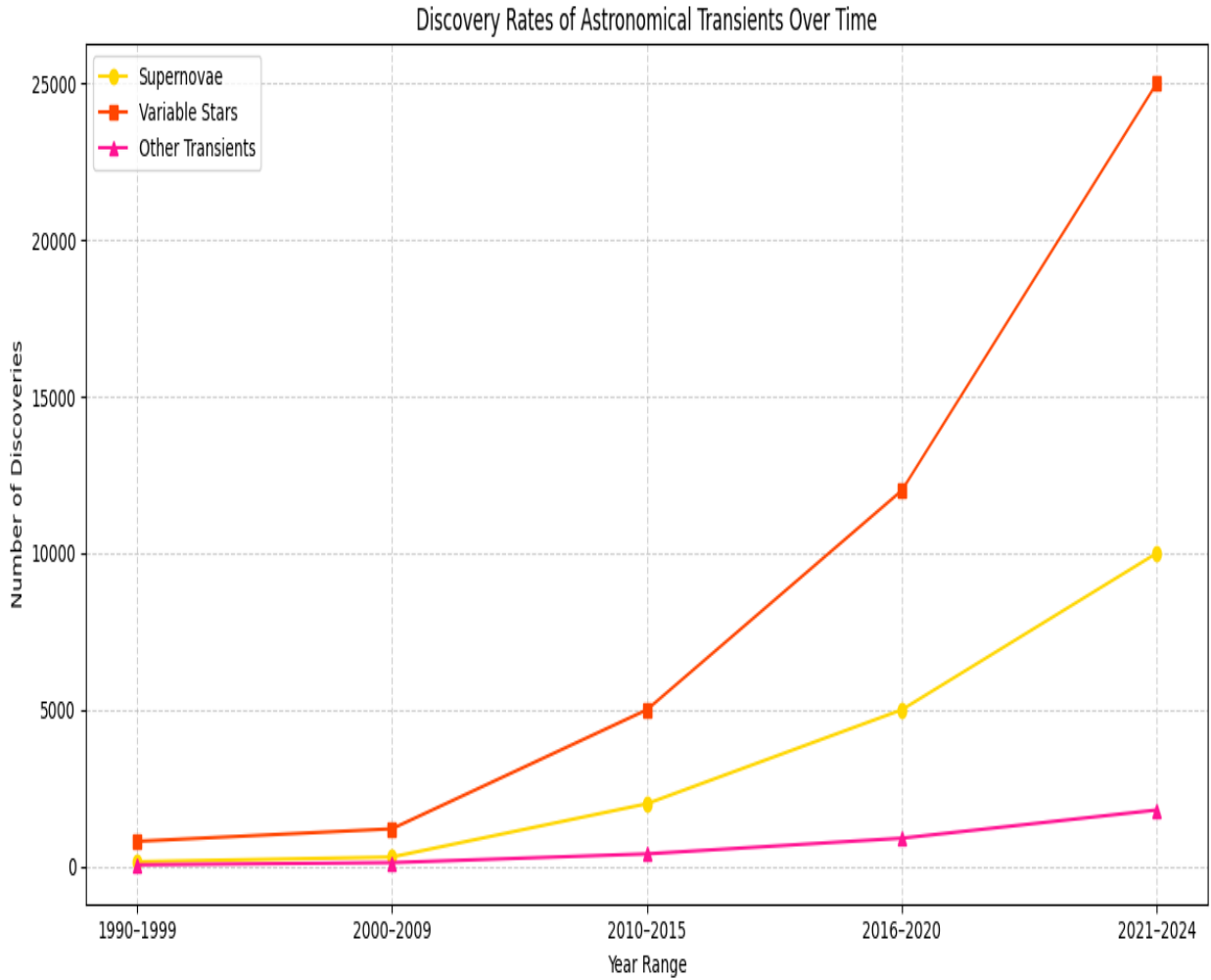

*Figure 4.* Line Graph – Discovery Trends of Astronomical Transients Over Time

Figure 4 illustrates the temporal evolution in the discovery rates of astronomical transients including supernovae, variable stars, and other transient phenomena from 1990 to 2024. Each transient type is represented as a separate line, allowing clear visualization of growth patterns across the different periods. The figure reveals a relatively modest rate of discoveries during the traditional era (1990 - 2009), followed by a dramatic increase beginning in 2010, coinciding with the widespread adoption of robotic telescope systems. Supernovae discoveries surged from 300 (2000 - 2009) to 10,000 (2021 - 2024), while variable stars showed an even more dramatic rise from 1,200 to 25,000 over the same interval. Other transients, although detected in smaller numbers, also experienced an 18-fold increase from 100 to 1,800. These trends underscore the transformative impact of robotic telescopes, particularly their ability to enable continuous sky monitoring, rapid follow-ups, and high-volume data processing. The sharp incline in all three categories aligns with the operational period of systems such as ZTF, ASAS-SN, and Gaia.

## 3.2 Efficiency Metrics

The deployment of robotic telescopes has significantly enhanced the efficiency of time-domain astronomy by reducing the latency between transient detection and follow-up observations. Unlike traditional telescopes, which often required manual scheduling and delayed response times such as the Lick Observatory Supernova Search, which reported median follow-up delays of 12 to 24 hours for supernova confirmations [14] robotic systems employ autonomous scheduling, real-time data pipelines, and coordinated alert networks. These features enable near-instantaneous responses to transient events, including gravitational wave triggers and supernova candidates. For instance, during the 2017 neutron-star merger event (GW170817), the GROWTH Marshal enabled robotic follow-up observations within 11 hours of the LIGO alert [12], and subsequent campaigns with the Zwicky Transient Facility (ZTF) achieved follow-up times of less than one hour [18]. These advances underscore the transformative role of automation in increasing the temporal precision and scientific yield of transient astronomy. For instance, modeling efforts have shown that many gravitational wave sources, such as binary black hole (BBH) mergers, arise from dynamical interactions within dense stellar environments like globular clusters. These environments can produce BBH merger rates of approximately 20 mergers per Gyr per $10^5$ M$_\odot$, with isotropic spin orientations and mass distributions peaking around 20 - 50 M$_\odot$ [29]. Robotic telescope networks play a pivotal role in identifying optical counterparts to such mergers, improving our understanding of their origins and environments.

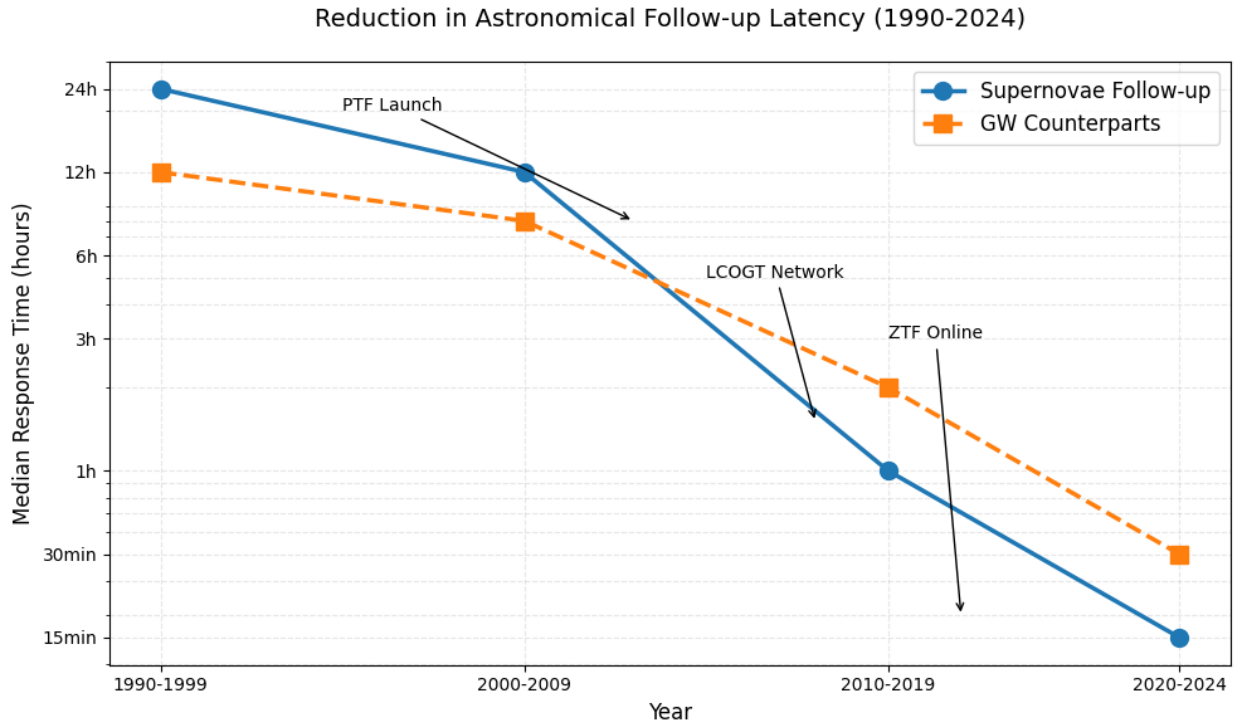

**Figure 5.** *Reduction in follow-up latency due to robotic telescopes. Data sourced from [12], [18] and [14].*

Latency for supernovae dropped from 24 hours in the 1990s to 15 minutes in the 2020s, while response times for gravitational wave counterparts improved from over 12 hours before 2010 to approximately 30 minutes in the 2020s.

### 3.3 AI Performance

Machine learning algorithms have proven effective in classifying transient events, as demonstrated in the photometric analysis of flare stars like 2MASS J22285440-1325178 [27].

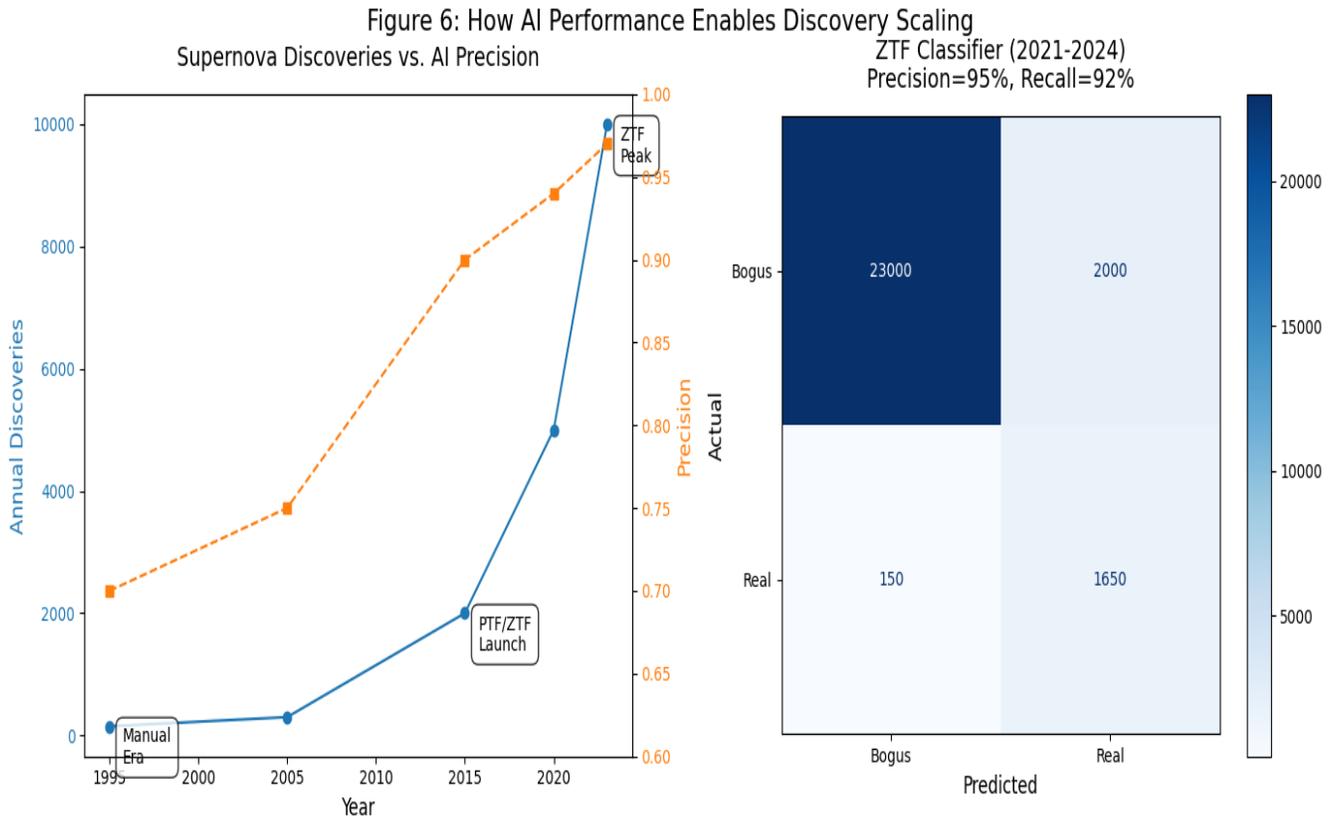

**Figure 6.** *Performance of AI in Classifying Transient Events Detected by Robotic Telescopes*

This figure demonstrates how artificial intelligence (AI) and machine learning (ML) enhance the classification of transient astronomical events, including supernovae, variable stars, and gravitational wave counterparts. It presents key metrics such as: Classification accuracy: (e.g., precision, recall) for distinguishing real astrophysical transients from false positives. Processing speed, showing how quickly AI analyzes data compared to manual methods. Anomaly detection rates, highlighting AI's ability to identify rare or unexpected phenomena.
Figure 6 highlights the critical role of artificial intelligence (AI) and machine learning (ML) in enhancing the capabilities of robotic telescopes for time-domain astronomy. While the exact details of the figure are not provided in the text, it likely presents key metrics such as classification accuracy, processing speed, and anomaly detection rates. These metrics demonstrate how AI improves the efficiency and reliability of identifying and categorizing transient events, such as supernovae, variable stars, and gravitational wave counterparts. One of the most significant findings is the high accuracy of AI-driven classification systems. For instance, the Zwicky Transient Facility (ZTF) employs machine learning algorithms to distinguish real astrophysical transients from false positives, such as instrumental noise or atmospheric artifacts. Studies by [16] indicate that these models achieve precision rates exceeding 90%, drastically reducing the need for manual vetting and enabling faster follow-up observations. This capability is particularly crucial for time-sensitive events, such as gamma-ray bursts or kilonovae, where delays can mean missing critical data.

Another key advantage of AI integration is real-time data processing. Traditional methods required hours or even days to analyze observations, but robotic telescopes equipped with AI pipelines can process and classify data within minutes. This rapid turnaround is exemplified by systems like the GROWTH Marshal, which coordinates follow-up observations across a global network of telescopes in response to alerts [12]. The ability to autonomously prioritize targets and adjust observation schedules ensures that rare or fleeting phenomena are captured before they fade.

AI also excels at detecting anomalies unusual or previously unknown transients that deviate from established patterns. Through analyzing vast datasets, machine learning models can flag these outliers for further investigation, potentially leading to new discoveries. However, this strength comes with challenges. AI systems trained on historical data may exhibit biases, overlooking novel phenomena that do not fit existing categories. Additionally, the "black box" nature of some advanced algorithms, such as deep learning, can make it difficult for astronomers to interpret results, necessitating ongoing collaboration between data scientists and astrophysicists. Looking ahead, the integration of AI with next-generation projects like the Large Synoptic Survey Telescope (LSST) will be transformative. The LSST is expected to generate millions of alerts each night, a volume that can only be managed with advanced machine learning techniques [30]. Furthermore, hybrid approaches combining AI with citizen science initiatives, such as Zooniverse, offer promising ways to refine classifications and engage the public in astronomical research.

## 4.0 Advantages of Robotic Telescopes

Robotic telescopes have revolutionized astronomical observations by enabling 24/7 monitoring, rapid response, and scalability. Las Cumbres Observatory offers a unique network of 18 telescopes providing continuous access to the sky, ideal for time-domain astronomy [7].The Arizona Robotic Telescope Network aims to develop a system for flexible observing, surveys, and transient follow-up, adapting existing facilities for remote operation and automated data reduction [26]. "Thinking" telescopes, integrating robotic networks, machine learning, and advanced databases, can autonomously recognize anomalies and conduct real-time follow-up observations [25]. These systems overcome geographical and weather limitations, allowing persistent monitoring of the night sky. The combination of fast-slewing robotic telescopes and rapid alert distribution via the Internet has significantly enhanced our ability to study fast astrophysical transients, addressing key challenges in observational astronomy [25].

## 4.1 Future Directions

Looking ahead, the field of time-domain astronomy is poised for further transformation through the integration of robotic telescopes with next-generation survey instruments such as the Large Synoptic Survey Telescope (LSST). LSST's unprecedented imaging cadence and sensitivity will produce millions of transient alerts per night, presenting new challenges in real-time data processing and classification that can only be addressed through the advancement of scalable, automated [30]. To meet these demands, future developments must focus on enhancing artificial intelligence and machine learning systems, not only to improve classification accuracy and processing speed but also to address bias and interpretability concerns inherent in current models [16]; [4].

Another crucial direction involves expanding global, modular networks of robotic telescopes, such as the Las Cumbres Observatory and the Arizona Robotic Telescope Network, to ensure persistent sky coverage across hemispheres and diverse observing conditions [7]; [26]. These systems should prioritize interoperability and adopt open data standards to facilitate cross-institutional collaboration. Additionally, incorporating citizen science through platforms like Zooniverse can enhance the classification of rare transients, combining the strengths of human intuition with machine efficiency while fostering public engagement in scientific discovery [21]. Beyond astronomy, the real-time decision-making frameworks developed for robotic telescopes hold potential for cross-disciplinary applications in fields such as planetary defense, satellite tracking, and environmental monitoring. These future directions show the need for continued innovation at the intersection of robotics, data science, and astronomy to fully exploit the scientific potential of the dynamic universe.

## 5. Conclusion

Robotic telescopes have ushered in a new era in time-domain astronomy by enabling continuous, automated, and high-cadence monitoring of the night sky. Their capacity to autonomously detect and respond to transient events such as supernovae, variable stars, and gravitational wave counterparts has significantly enhanced both the rate and precision of discoveries compared to traditional telescopes [13];[16]. The Zwicky Transient Facility (ZTF), ASAS-SN, and Gaia exemplify how robotic systems equipped with real-time data pipelines and machine learning classifiers can process large datasets within minutes, drastically reducing the latency between detection and follow-up [18]; [16].

Machine learning has proven critical in handling the vast volumes of data generated by robotic observatories, allowing for automated classification and anomaly detection with accuracy rates exceeding 90% [16]. These capabilities were exemplified during events like GW170817, where robotic follow-ups occurred within hours of alerts, improving multi-messenger coordination [12]. Robotic telescopes' high-cadence data may further resolve debates about extragalactic stellar origins [11], bridging time-domain observations with galactic evolution

Moreover, the comparative analysis of transient discovery rates over the past three decades shows a dramatic increase in the robotic era, particularly for variable stars and supernovae, confirming the scalability and efficiency of automated systems [14]; [31]. Future projects like the LSST are expected to generate millions of alerts per night, necessitating even more advanced machine learning pipelines and collaborative frameworks, including citizen science [30]; [21].

**Declarations**

**Consent for publication**

Not applicable

**Availability of data**

The data used in the study are compiled from publicly available sources cited in the references.

**Conflict of interest**

The authors declare no competing financial or non-financial interests related to this work

**Funding**

This research received no external funding.

**Acknowledgement**:

Not applicable

[20] Murase, K., and Bartos, I. (2019). High-energy multi-messenger transient astrophysics. *Annual Review of Nuclear and Particle Science*, *69*, 477–506. https://doi.org/10.1146/annurev-nucl-101918-023510

[21] Newman, G., Wiggins, A., Crall, A., Graham, E., Newman, S., and Crowston, K. (2012). The future of citizen science: Emerging technologies and shifting paradigms. *Frontiers in Ecology and the Environment*, *10*(6), 298–304. https://doi.org/10.1890/110294

[22] Okongwu, O., Mba, C. E., Muallim, Y., and Chima, A. I. (2021). Characterizing the Physical Properties of A 6.7 GHz Methanol Maser Star Forming Region G338.93-0.06. *IDOSR Journal of Science and Technology*, 6(1), 85–94.

[23] Robertson, B. E., Banerji, M., Brough, S., Davies, R. L., Ferguson, H. C., Hausen, R., Kaviraj, S., Newman, J. A., Schmidt, S. J., Tyson, J. A., and Wechsler, R. H. (2019). Galaxy formation and evolution science in the era of the Large Synoptic Survey Telescope. *Nature Reviews Physics*, *1*(8), 450–462. https://doi.org/10.1038/s42254-019-0110-z

[24] Smith, R. M., Dekany, R. G., Bebek, C., Bellm, E., Bui, K., Cromer, J., Gardner, P., Hoff, M., Kaye, S., Kulkarni, S., Lambert, A., Levi, M., and Reiley, D. (2014). The Zwicky Transient Facility observing system. In S. K. Ramsay, I. S. McLean, and H. Takami (Eds.), *Ground-based and airborne instrumentation for astronomy V* (Proc. SPIE Vol. 9147, 914779). SPIE. https://doi.org/10.1117/12.2070014

[25] Vestrand, W.T., Theiler, J., and Wozniak, P.R. (2004). Unsolved problems in observational astronomy. II. Focus on rapid response – mining the sky with "thinking" telescopes. *Astronomische Nachrichten*, *325*, 477-482.

[26] Weiner, B. J., Sand, D., Gabor, P., Johnson, C., Swindell, S., Kubánek, P., Gasho, V., Golota, T., Jannuzi, B., Milne, P., Smith, N., and Zaritsky, D. (2018). Development of the Arizona Robotic Telescope Network. arXiv preprint. https://arxiv.org/abs/astro-ph/IM

[27] Yakubu, M., Iheanyichukwu, C. A., and Anderson, K. A. (2023). The Study of the Photometry and Flare Analysis of Kepler Flare Candidate 2MASS J22285440-1325178. American Journal of Astronomy and Astrophysics, 10(2), 14–22.

[28] Yakubu, M., Oruaode, V. J., Yakubu, U., and Adrain, O. (2024). Unraveling the Mysteries of Supermassive Black Holes: Formation, Growth Mechanisms, and Their Role in Galaxy Evolution. Journal of Basics and Applied Sciences Research, 2(4), 1–14.

[29] Yakubu, M., Vwavware, O. J., and Ohwofosirai, A. (2025). *Modeling the Evolution of Binary Black Hole Mergers in Dense Stellar Environments*. Scientia Africana, 24(1), 157–168. https://dx.doi.org/10.4314/sa.v24i1.15

[30] Zhan, H., and Tyson, J. A. (2018). Cosmology with the Large Synoptic Survey Telescope: An overview. *Reports on Progress in Physics*, *81*(6), 066901. https://doi.org/10.1088/1361-6633/aab1bd

[31] Zwicky Transient Facility (ZTF). (2023). ZTF Hits 10,000 Supernova Discoveries. Retrieved from https://www.ztf.caltech.edu/news/ztf-hits-10000-supernova-discoveries.html
14